\documentclass[
    ,final            
  ]
  {aipproc}

\layoutstyle{6x9}
\usepackage{graphicx}

\newcommand{\De}{\Delta}
\newcommand{\Sg}{\Sigma}
\newcommand{\Sgs}{\Sigma^*}
\newcommand{\X}{\Xi^*}

\newcommand{\be}{\begin{equation}}
\newcommand{\ee}{\end{equation}}
\newcommand{\Ld}{\Lambda(1520)}
\begin{document}

\title{Chiral dynamics of the $\Ld$ in coupled channels tested in the
$K^-p\to\pi\pi\Lambda$ reaction}


\author{Sourav~Sarkar\footnote{Present address:Variable Energy Cyclotron Centre,
1/AF, Bidhannagar, Kolkata-700091, India}, L.~Roca, E.~Oset, V.~K.~Magas and M.~J.~Vicente~Vacas}{
  address={Departamento de F\'{\i}sica Te\'orica and IFIC,
Centro Mixto Universidad de Valencia-CSIC, 
Investigaci\'on de Paterna, Aptdo. 22085, 46071 Valencia, Spain}
}

\begin{abstract}
The $\Ld$ resonance is generated dynamically in a unitary coupled channel
framework with the $\pi\Sgs$ and $K\X$ channels in $s$-wave and $\pi\Sg$ and
$\bar K N$ channels in $d$-wave. The dynamics of this resonance close to and
above threshold is then tested through the reactions $K^-p\to\pi\pi\Lambda$ and
a good agreement with the experimentally measured cross section is observed.
\end{abstract}

\maketitle

Application of unitary
techniques to the lowest order chiral Lagrangian involving the octet of pseudoscalar
mesons and the decuplet of baryons have led to the successful 
generation of a number of $\frac{3}{2}^-$ resonances~\cite{lutz,Sarkar:2004jh,VicenteVacas:2005dn}. From the information of
the pole positions and couplings to the channels involved of these
resonances could be associated to the $N^*(1520)$, $\De(1700)$, $\Lambda(1520)$, $\Sigma(1670)$,
$\Sigma(1940)$, $\Xi(1820)$ tabulated by the Particle Data Group (PDG). 
The $\Ld$, in particular, is generated dynamically in the coupled channel interaction of the $\pi\Sgs$ and
$K\X$ channels and appears at a higher energy than the nominal one and with a
width much larger than the physical width~\cite{Sarkar:2004jh}. Since the width of the $\Lambda(1520)$ resonance comes basically from the
decay into
the $\bar{K} N$ and $\pi \Sigma(1193)$, the introduction of these channels is 
mandatory to reproduce the shape of the $\Lambda(1520)$ resonance. The novelty with
respect to the other channels already accounted for in~\cite{Sarkar:2004jh},
 which couple in $s$-wave,
 is that these new channels couple in $d$-waves. These channels are  
introduced phenomenologically  
using for the vertices 
$\bar{K}N\to\bar{K}N$, $\bar{K}N\to\pi\Sigma$ and
$\pi\Sigma\to\pi\Sigma$ effective
transition potentials which are 
proportional to the incoming and outgoing momentum squared. 
 Denoting  $\pi\Sigma^*$, $K\Xi^*$,  $\bar K N$  and
$\pi\Sigma$ channels by $1$, $2$, $3$ and $4$ respectively, the
matrix containing the tree level amplitudes is written 
as~\cite{Sarkar:2005ap,luis_prep}

\renewcommand{\arraystretch}{1.2}
\be
V=\left( 
\begin{array}{cccc}
C_{11}(k_1^0+k_1^0)\ & C_{12}(k_1^0+k_2^0) & \gamma_{13}\,q_3^2 &\gamma_{14}\,q^2_4 \\
C_{21}(k_2^0+k_1^0)\ & C_{22}(k_2^0+k_2^0) & 0 & 0 \\
\gamma_{13}\,q_3^2 & 0 & \gamma_{33}\, q^4_3 & \gamma_{34} \,q_3^2 \,q^2_4\\
\gamma_{14}\,q^2_4  & 0 & \gamma_{34} \,q_3^2 \,q_4^2 &  \gamma_{44}\, q^4_4
\end{array}
\right)~,
\ee
\noindent
where $q_i=\frac{1}{2\sqrt{s}}\sqrt{[s-(M_i+m_i)^2][s-(M_i-m_i)^2]}$,
$k_i^0=\frac{s-M_i^2+m_i^2}{2\sqrt{s}}$ 
and $M_i(m_i)$ is the baryon(meson) mass. 
The coefficients $C_{ij}$ are $C_{11}=\frac{-1}{f^2}$,
 $C_{21}=C_{12}=\frac{\sqrt{6}}{4f^2}$ and $C_{22}=\frac{-3}{4f^2}$,
where $f$ is $1.15f_\pi$, with $f_\pi$ ($=93$ MeV) the pion decay constant,
which is taken as an average between $f_\pi$ and
$f_K$. The elements $V_{11}$, $V_{12}$, $V_{21}$, $V_{22}$ come from the
lowest order chiral Lagrangian involving the decuplet of baryons and
the octet of pseudoscalar mesons~\cite{Sarkar:2004jh,lutz}.
We neglect the elements $V_{23}$ and $V_{24}$ which involve the
tree level interaction of the $K\Xi^*$ channel to the $d$-wave channels
because the $K\Xi^*$ threshold is far away from the $\Ld$.
We also emphasize that the consideration of the width of the $\Sigma^*$ resonance
in the loop function $G$ is crucial in order to account properly for
the $\pi\Sigma^*$ channel since the threshold lies in the $\Lambda(1520)$
region. This is achieved through the convolution of the $\pi\Sigma^*$ loop
function with the spectral distribution considering the $\Sigma^*$
width.

In the model described so far we have as unknown parameters
$\gamma_{13}$, $\gamma_{14}$, $\gamma_{33}$, $\gamma_{34}$,
$\gamma_{44}$  in the $V$ matrix. Apart from these, there is also
the freedom in the value of the subtraction constants in the loop
functions. We will consider one subtraction constant for the
$s$-wave channels ($a_0$) and one for the $d$-wave ones ($a_2$).
Despite the apparent large number of free parameters in the $V$
matrix, it is worth emphasizing that the largest matrix elements
are $V_{11}$, $V_{12}$ and $V_{22}$ \cite{Sarkar:2004jh}
which come from a chiral  Lagrangian  without any
free parameters. Due to the $d$-wave behavior
the other ones are much  
smaller, as expected, as we see from the values of the parameters $\gamma$ given below.
In order to obtain these parameters we fit 
the partial wave amplitudes obtained by using the $V$ matrix given above
as the kernel in
the Bethe-Salpeter equation
to the
experimental results on the $\bar K N$ and $\pi\Sigma$ scattering
amplitudes in $d$-wave and $I=0$.
We use experimental data from~\cite{Gopal:1976gs,Alston-Garnjost:1977rs} where 
$\bar K N\to\bar K N$ and $\bar K N\to\pi\Sigma$ amplitudes are
provided from partial wave analysis.
From the fit we obtain the subtraction constants $a_0=-1.8$ for the $s$-wave
channels and $a_2=-8.1$  for the $d$-wave channels. The unknown constants in the
$V$ matrix are given by $\gamma_{13}=0.98$ and $\gamma_{14}=1.1$ in units
of $10^{-7}$ MeV$^{-3}$ and $\gamma_{33}=-1.7$, $\gamma_{44}=-0.7$ and
$\gamma_{34}=-1.1$ in units of $10^{-12}$ MeV$^{-5}$. 


From the imaginary part of the amplitudes it is straightforward to
obtain the couplings of the $\Ld$ to the different channels. Up to a 
global sign of one of the couplings (we choose $g_1$ to be
positive), the couplings we obtain are shown in Table~\ref{tab:coup}.
\begin{table}[h]
\begin{tabular}{|c|c|c|c|}
\hline
$g_1$ &$g_2$ &$g_3$ &$g_4$ \\	
\hline
 $0.91$ & $-0.29$ & $-0.54$ & $-0.45$ \\
\hline
\end{tabular}
\caption{Couplings of the $\Ld$ resonance to the different channels}
\label{tab:coup}
\end{table}
 We can see from the values that the $\Ld$ resonance 
couples most strongly to the $\pi\Sgs$ channel.
The fact that we are able to predict the value of this coupling is a
non trivial consequence of the unitarization procedure that we
employ.

\begin{figure}[t]
\centerline{
\includegraphics[width=1.0\textwidth]{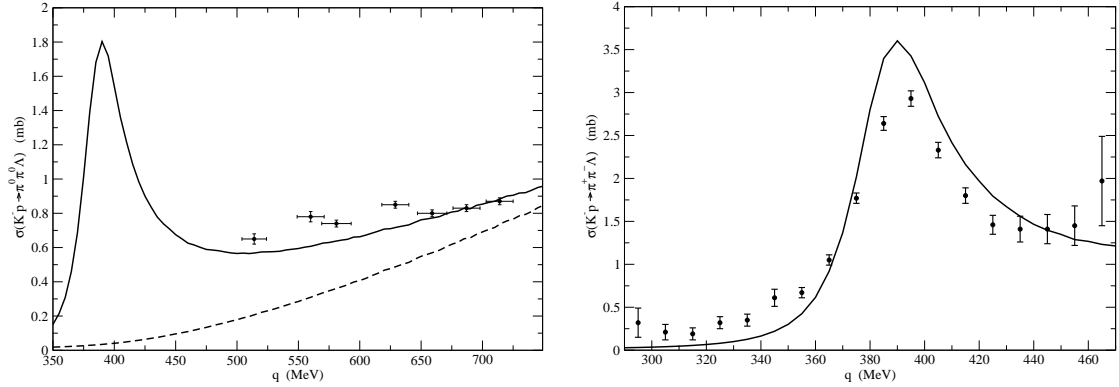}
}
\vspace*{0.5cm}
\caption{Result for the $K^-p\to\pi^0\pi^0\Lambda$ (left) and 
$K^-p\to\pi^+\pi^-\Lambda$ (right) cross section.
}
\label{fig:nef_mast}
\end{figure}

The prediction of the amplitudes involving $\pi\Sigma^*$ channels can
be checked in particular reactions where this channel could play an
important role. We evaluate the cross section for $K^-p\to\pi\pi\Lambda$ in the
lines of~\cite{Sarkar:2005ap} but using the new coupled channel
formalism. The mechanisms and the 
expressions
 for the amplitudes and
the cross sections can be found in~\cite{Sarkar:2005ap} where,
apart from the coupled channel unitarized amplitude, other
mechanisms of relevance above the $\Ld$ peak were also included.
In fig.~\ref{fig:nef_mast} 
 we show our results for 
$K^-p\to\pi^0\pi^0\Lambda$ and $K^-p\to\pi^+\pi^-\Lambda$
 cross section on the left and right panels along with experimental data
from refs.~\cite{Prakhov:2004ri} and ~\cite{Mast:1973gb} respectively.
The dashed line in the left figure  represents the contribution from mechanisms 
 other than the unitarized
coupled channels, and the solid line gives the coherent sum of all
the processes.
These cross sections depend essentially on the amplitude $T_{\bar K
N\to\pi\Sgs}$ which we obtain from our 
analysis.

In conclusion, we have done a unitary coupled channel analysis of the $\Ld$ resonance
using the  $\pi\Sigma^*$, $K\Xi^*$ channels in $s$-wave and the $\bar{K}N$ and
$\pi\Sigma$ channels in $d$-wave. Our predictions of the amplitudes and
couplings of the $\Ld$ to the different channels were 
tested in the
 $K^-p\to\Lambda\pi^0\pi^0$ and $K^-p\to\Lambda\pi^+\pi^-$
reactions for which
the magnitude of the absolute cross section agrees fairly well
with experimental data at energies 
close to and slightly above the $\Ld$ region.


\begin{theacknowledgments}
This work is partly supported by DGICYT contract number BFM2003-00856,
and the E.U. EURIDICE network contract no. HPRN-CT-2002-00311.
This research is part of the EU Integrated Infrastructure Initiative
Hadron Physics Project under contract number RII3-CT-2004-506078. 
\end{theacknowledgments}

\end{document}